# Improved lattice Boltzmann method for conjugate magnetohydrodynamic simulations

Jun Li, Wai Hong Ronald Chan, Kun Ting Eddie Chua

Institute of Advanced Intelligence and Computing (IAIC), Agency for Science, Technology and Research (A*STAR), 1 Fusionopolis Way, #16-16 Connexis North Tower, Singapore 138632, Singapore

**Abstract:** In the simulation of conjugate magnetohydrodynamic (MHD) flows, where a moving conducting fluid dynamically interacts with bounding conducting walls under the influence of an externally imposed magnetic field, resolving the electrical conductivity transition across the fluid-solid interface introduces significant numerical challenges. In problems with high magnetic Reynolds numbers, it is required to solve the full magnetic induction formulation that has a curl-of-curl term for the magnetic diffusion, which is simplified to a divergence term in the original vector-valued lattice Boltzmann method (LBM). The resulting LBM scheme is valid for simulating fluid domains with a constant conductivity where its bounding solid walls are modelled using appropriate boundary conditions. The current study shows that this scheme is also valid for two-dimensional conjugate MHD simulations where the magnetic field component with non-zero gradients is perpendicular to the conductivity gradient. For general conjugate MHD simulations, we improve the LBM scheme by considering the difference between the curl-of-curl term and the simplified divergence form, which is treated as a source term in the LBM evolution algorithm and computed using an efficient LBM discretisation scheme. The improved LBM scheme automatically satisfies the required conjugate constraints at the fluid-solid interface as a volume integral inside two adjacent boundary layers. The accuracy of the improved LBM scheme is verified by piecewise analytical solutions in benchmark problems with an abrupt conductivity jump across the fluid-solid interface.

## 1. Introduction

The study of magnetohydrodynamic (MHD) flows is important for understanding solar coronal loop dynamics [1], planetary core-mantle boundary interactions [2], electromagnetic pumping [3], electromagnetic braking for stable casting [4], liquid-metal cooling [5, 6, 7] and tritium breeding blankets [8], as studied in astrophysics, geophysics, advanced manufacturing and nuclear fusion, respectively. In problems with a variable electrical conductivity across fluid-solid interfaces, solving the full magnetic induction formulation remains a numerical challenge. This full MHD formulation contains a curl-of-curl diffusion term that is required to capture solutions with sharp structural transitions. Even though the alternative electric potential formulation is valid in applications with small magnetic Reynolds numbers [4, 5], it could still be complicated when variable electrical conductivities are considered for conjugate simulations.

MHD flows can be modelled by coupling the magnetic induction equation for the magnetic field $\vec{B}$ and the incompressible Navier–Stokes equations for the flow velocity $\vec{u}$ and hydrodynamic pressure $p$ fields. The magnetic induction formulation includes the continuity equation

$$\nabla \cdot \vec{u} = 0, \tag{1}$$

the momentum equation

$$\rho_0 \left( \frac{\partial \vec{u}}{\partial \mathrm{t}} + \vec{u} \cdot \nabla \vec{u} \right) = -\nabla p + \rho_0 \nu \Delta \vec{u} + \rho_0 \vec{a} + \vec{J} \times \vec{B}, \tag{2}$$

and the full magnetic induction equation:

$$\frac{\partial \vec{B}}{\partial \mathrm{t}} = \nabla \times \left( \vec{u} \times \vec{B} \right) - \nabla \times (\eta \nabla \times \vec{B}), \tag{3}$$

which is solved together with the divergence-free condition for the magnetic field

$$\nabla \cdot \vec{B} = 0, \tag{4}$$

where $\rho_0$ is the fluid mass density, $t$ is time, $\nu$ is the kinematic viscosity, $\vec{a}$ is the external body force per unit mass, $\eta = 1/(\mu\sigma)$ is the magnetic diffusivity, $\sigma$ is the electrical conductivity, $\mu$ is the magnetic permeability, $\vec{J}$ is the induced electric current and computed as $\mu^{-1}\nabla \times \vec{B}$, and $\vec{J} \times \vec{B}$ is the magnetic component of the Lorentz force per unit volume. The parameter $\sigma$ equals $\sigma_{\mathrm{f}}$ and $\sigma_{\mathrm{s}}$ inside the fluid and solid domains, respectively, making $\eta$ spatially discontinuous, and $\vec{u}$ is zero inside the solid domain in conjugate simulations.

The above magnetic induction formulation of MHD flows can be efficiently solved in the fluid domain using a vector-valued LBM, as detailed in the seminal work by Dellar [9]. It has been successfully applied to simulate MHD turbulence [10] and MHD flows in porous media [6]. To the best of our knowledge, the only fluid-solid conjugate LBM simulation of MHD flows is found in one of several studied cases of Ref. [8], where the curl-of-curl diffusion term $-\nabla \times (\eta \nabla \times \vec{B})$ is simplified to a divergence term $\nabla \cdot (\eta \nabla \vec{B})$, as assumed by the adopted LBM scheme [9]. Other reported simulations are conducted only inside the fluid phase and treat the bounding wall using appropriate boundary conditions. In the current study, we address the challenge of solving the full magnetic induction formulation for general applications and propose improvements to the existing LBM algorithm [9, 10, 11, 12] to account for the difference between the curl-of-curl diffusion term and the simplified divergence term due to variable electrical conductivities.

## 2. Improved LBM algorithm for conjugate MHD simulations

The original LBM algorithm proposed in Ref. [9] replaces $-\nabla \times (\eta \nabla \times \vec{B})$ of Eq. (3) by a divergence term $\nabla \cdot (\eta \nabla \vec{B})$, which is valid only when $\eta$ is constant, since the latter can be directly recovered by the Chapman–Enskog expansion. Therefore, the governing equations (1)–(4) for problems with a constant $\eta$ can be solved in two-dimensional (2-D) and three-dimensional (3-D) LBM simulations, as detailed in Refs. [9, 10, 11, 12]. In the current study, we improve the LBM algorithm by including the difference between the two operators as a source term $\vec{S}$, which is required for variable $\eta$, namely by solving Eq. (3) in the following form:

$$\frac{\partial \vec{B}}{\partial \mathrm{t}} = \nabla \times \left(\vec{u} \times \vec{B}\right) + \nabla \cdot \left(\eta \nabla \vec{B}\right) + \vec{S}, \tag{5}$$

where in our notation, $\vec{S} = -\nabla\vec{B} \cdot \nabla\eta$ has components $S_{i\in\{1,2,3\}} = -\sum_{j=1}^{3} \frac{\partial B_j}{\partial x_i}\frac{\partial \eta}{\partial x_j}$ for the three Cartesian coordinate directions. Correspondingly, the improved LBM algorithm for solving Eq. (5) is:

$$\vec{g}_\alpha(\vec{x} + \vec{e}_\alpha \Delta t, t + \Delta t) = \vec{g}_\alpha(\vec{x}, t) + \frac{\vec{g}_\alpha^{\mathrm{eq}}(\vec{x},t) - \vec{g}_\alpha(\vec{x},t)}{\tau_\eta} + \omega_\alpha \Delta t \vec{S}, \tag{6}$$

where $\vec{g}_\alpha$ is the vector-valued distribution function and its equilibrium distribution is:

$$\vec{g}_\alpha^{\mathrm{eq}}(\vec{B},\vec{u}) = \omega_\alpha[\vec{B} + \frac{4}{c^2}\vec{e}_\alpha \cdot (\vec{u}\vec{B} - \vec{B}\vec{u})], \tag{7}$$

where the lattice velocities $\vec{e}_\alpha$ and weight coefficients $\omega_\alpha$ are selected according to the D3Q7 lattice model for 3-D cases, $\vec{x}$ is the spatial coordinates, $\Delta t$ is the timestep, $\vec{B} = \sum_\alpha \vec{g}_\alpha + 0.5\Delta t\vec{S}$ is computed as a summation of $\vec{g}_\alpha$ over $\alpha \in [0, Q-1]$ with a correction for $\vec{S}$, and the relaxation time $\tau_\eta$ is computed as $\tau_\eta = 0.5 + 4\eta/(c^2\Delta t)$ using $c = \Delta x/\Delta t$ and the uniform grid size $\Delta x$. Note that the first-order moment of $\vec{g}_\alpha^{\mathrm{eq}}$ is equal to an antisymmetric tensor $\vec{u}\vec{B} - \vec{B}\vec{u}$ [9], corresponding to $\nabla \cdot (\vec{u}\vec{B} - \vec{B}\vec{u})$ on the left-hand size of the recovered governing equation for $\vec{B}$ that is equivalent to $\nabla \times \left(\vec{u} \times \vec{B}\right)$ on the right-hand side of Eq. (5).

The spatial gradient of $\eta$ in the source term $\vec{S}$ is computed as:

$$\nabla\eta(\vec{x}) = \frac{3}{c^2\Delta t}\sum_\alpha \omega_\alpha \vec{e}_\alpha \eta(\vec{x} + \vec{e}_\alpha \Delta t), \tag{8}$$

where $\vec{e}_\alpha$ and $\omega_\alpha$ are selected according to the D3Q19 lattice model for 3-D cases, which has more spatial differentiation stencils for better numerical accuracy and stability than the D3Q7 lattice model. Additionally, the tensor components of $\nabla\vec{B}$ in $\vec{S}$ can be conveniently computed from the local non-equilibrium distribution using the same D3Q7 model of Eq. (6) for $\vec{e}_\alpha$ and $\omega_\alpha$, as first proposed in Ref. [9] for computing $\vec{J} = \mu^{-1}\nabla \times \vec{B}$:

$$\frac{\partial B_j}{\partial x_i} = \frac{-4}{c^2\Delta t\tau_\eta}\sum_\alpha e_{\alpha,i}(g_{\alpha,j} - g_{\alpha,j}^{\mathrm{eq}}). \tag{9}$$

The LBM scheme for solving Eqs. (1) and (2) of the flow dynamics with the Lorentz force has been detailed in Refs. [9, 10, 11, 12]. The two LBM schemes are executed concurrently using a two-way coupling scheme for $\vec{u}$ and $\vec{B}$, respectively. The continuity equation of Eq. (1) is approximately satisfied by having small Mach numbers; similarly, the inherent divergence-free property of Eq. (4) holds at small dimensionless Alfvén velocities [8, 9] and has been verified in Refs. [9, 12]. A general derivation of various LBM boundary schemes has been detailed in our previous work [13] for multi-physics simulations, which considers the three conventional types of boundary conditions (i.e., Dirichlet, Neumann and Robin) at possibly moving and curved boundaries.

It can be proved that the improved LBM scheme with a source term $\vec{S}$ automatically imposes the required conjugate constraints at the fluid-solid interface as a volume integral inside two adjacent boundary layers, as discussed in the Appendix.

## 3. Simulation results

Our LBM simulations of MHD flows have been previously verified in rectangular channels [12], circular pipes [13], as well as in a cross-channel [6], where the magnetic induction equation is solved only inside the fluid domain and appropriate boundary conditions are imposed at the fluid-solid interface for different electrical conductivity ratios $C = \sigma_\mathrm{s}/\sigma_\mathrm{f}$. The current study is extended to conjugate simulations of the full magnetic induction equation where $\sigma_\mathrm{s}$ and $\sigma_\mathrm{f}$ take distinct values at adjacent solid and fluid grid points, respectively, without applying any smoothing techniques.

Two benchmark problems with analytical solutions are adopted to verify the proposed LBM scheme in two steps. In the first problem, conjugate MHD flows are simulated inside a circular pipe with a finite wall thickness where the solid-fluid coupling is present but the source term $\vec{S} = -\nabla\vec{B} \cdot \nabla\eta$ happens to be zero due to the 2-D nature of the solution, as applied in Refs. [8, 14], which is because the component $B_j$ with a non-zero gradient and the non-zero components $\partial\eta/\partial x_j$ are never aligned. In general 3-D problems, the presence of $\vec{S}$ acts as a cross-coupling mechanism: a gradient in one magnetic field component that is not perpendicular to the magnetic diffusivity gradient can generate or alter another component of the magnetic field. Note that the numerical challenges of conjugate MHD simulations stem from the abrupt jump in $\eta$ of the magnetic diffusion term, rather than the magnetic induction term due to the flow. Therefore, the second problem is designed to focus on $\vec{S}$ and sets the flow-induced term $\nabla \times (\vec{u} \times \vec{B})$ of Eq. (5) to zero using $\vec{u} \equiv \vec{0}$, which makes it easy to obtain analytical MHD conjugate solutions in 3-D cases. Therefore, these two test cases verify the capability of the proposed LBM scheme in solving all terms of Eq. (5).

### *3.1. Simulation of conjugate MHD flows inside a pipe*

In our previous study on MHD flows inside circular pipes [13], the magnetic field is simulated only inside the fluid domain and the solid domain with a finite wall thickness is modelled by the Shercliff boundary condition. In the current study, we explicitly simulate the solid domain for conjugate MHD flows. Variables and parameters with the standard physical (SI) units are used in our formulation and discussion. The parameters $\rho_0$, $\rho_0\nu$, $\sigma_\mathrm{f}$, and $\mu$ are set to unity for simplicity and the solid electrical conductivity $\sigma_\mathrm{s}$ is set by the ratio $C = \sigma_\mathrm{s}/\sigma_\mathrm{f}$. These parameters can take on non-unity values for practical configurations, but the obtained dimensionless quantities will remain unchanged, as shown in our previous study [6]. The circular pipe is set with an inner radius $R = 1$ and a wall thickness $\delta = 0.2R$, which are discretised using uniform grid points with a grid size $\Delta x = 0.04$. After fixing $\Delta x$, the timestep $\Delta t$ is set using $c = \Delta x/\Delta t = 250$, and the three LBM relaxation times are determined from $\nu$, $\eta_\mathrm{f} = 1/(\mu\sigma_\mathrm{f})$ and $\eta_\mathrm{s} = 1/(\mu\sigma_\mathrm{s})$, respectively.

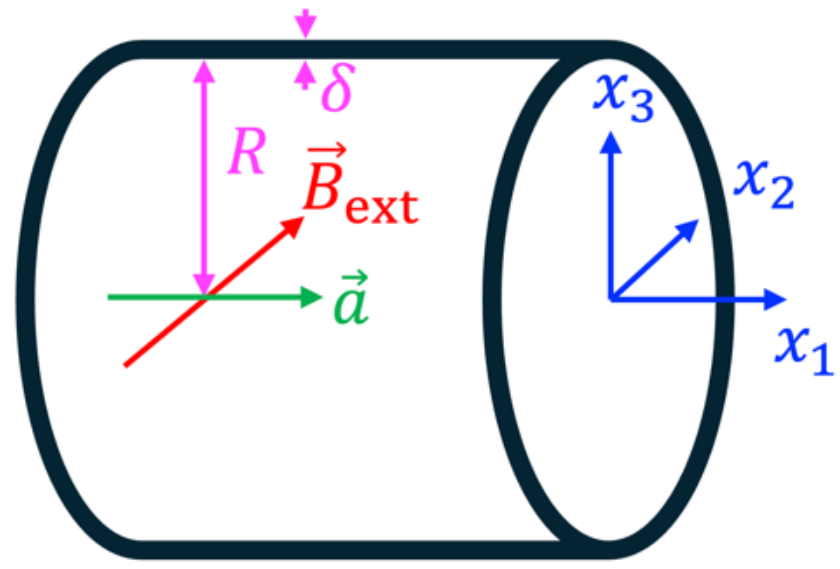

Fig. 1: Schematic for conjugate MHD flows inside a pipe with walls of finite thickness and conductivity.

As shown in Fig. 1, the flow is driven using an external body force $a_1 = 1$ in the $x_1$ direction along the pipe axis. The external magnetic field $\vec{B}_{\text{ext}}$ is imposed in the $x_2$ direction and the Hartmann number is defined as $Ha = |B|_{\text{ext}} R\sqrt{\sigma_{\text{f}}/(\rho_0 \nu)}$ using the magnitude $|B|_{\text{ext}}$. The simulation results are normalised as follows for analysis:

$$B_1^* = \frac{B_1}{U_1 \mu \sqrt{\rho_0 \nu \sigma_{\text{f}}}}, \qquad u_1^* = \frac{u_1}{U_1}, \qquad x_i^* = \frac{x_i}{R}, \tag{10}$$

where $U_1$ is the mean flow velocity of $u_1$, and $x_i$ with $i = 2$ and 3 are coordinates in the two transverse directions. Periodic boundary conditions are applied in the $x_1$ direction, the insulating boundary condition with $\vec{B} = \vec{B}_{\text{ext}}$ is imposed on the external boundary of the pipe, and the no-slip boundary condition for $\vec{u}$ is applied to the internal boundary.

Note that the contour lines of $B_1$ are the electric current traces in 2-D cases. As shown in Fig. 2(a), (c) and (e), the current is induced in the $x_3$ direction due to flow and some currents circulate within the fluid domain enveloped by the dashed blue line, while others are closed through the solid wall. The currents flowing across the interface into the solid wall increase with $C$ and the magnitude of $J_3 = (-\partial B_1/\partial x_2)/\mu$ increases as well. Correspondingly, the velocity magnitude inside the pipe of Fig. 2(b), (d) and (f) is decreased due to the magnitude increase of the resisting Lorentz force $|-J_3 B_{\text{ext},2}|$. The comparisons in Fig. 2 shows that the LBM results at steady state agree very well with the piecewise analytical solutions [15] in the studied cases.

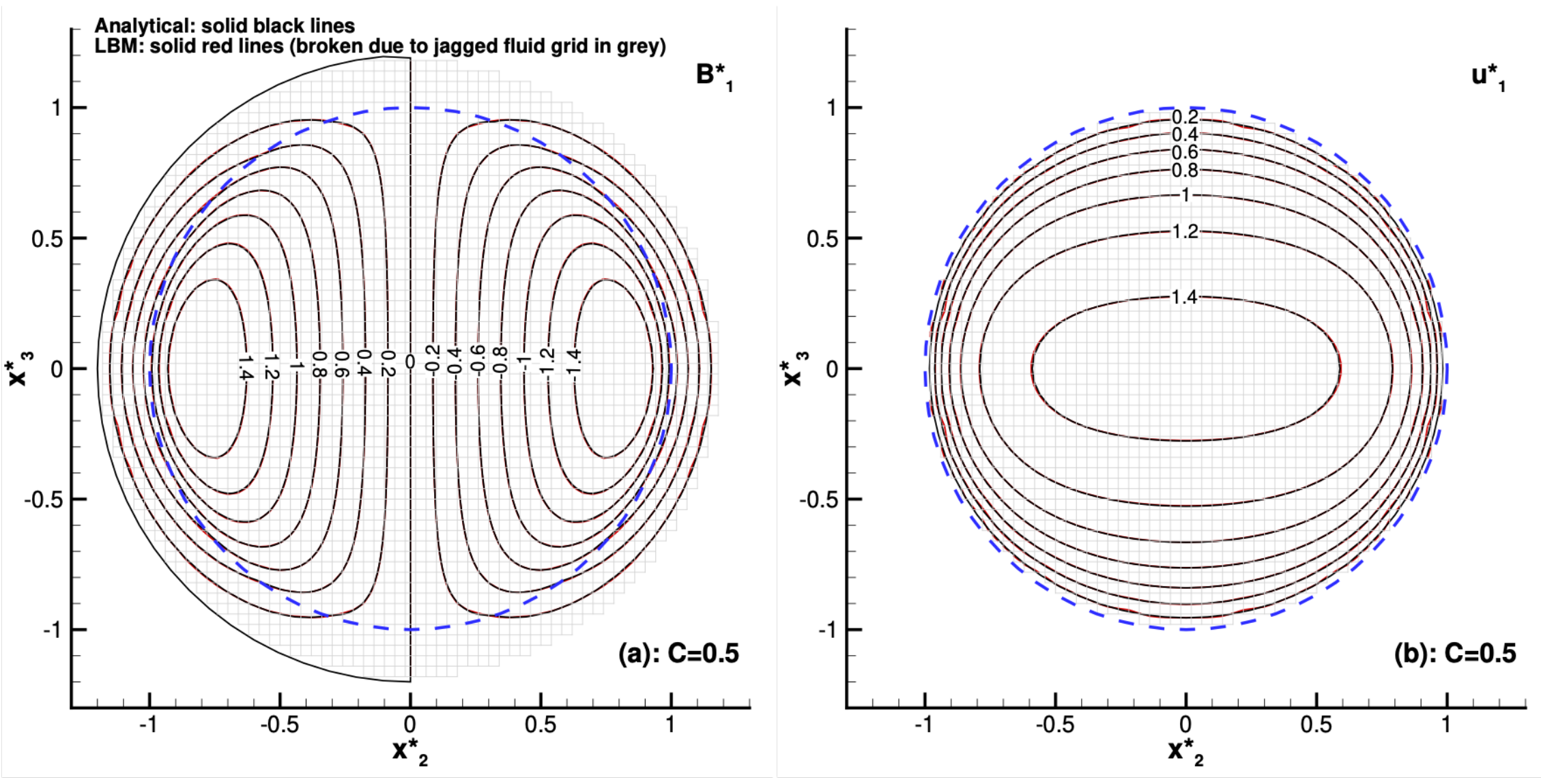

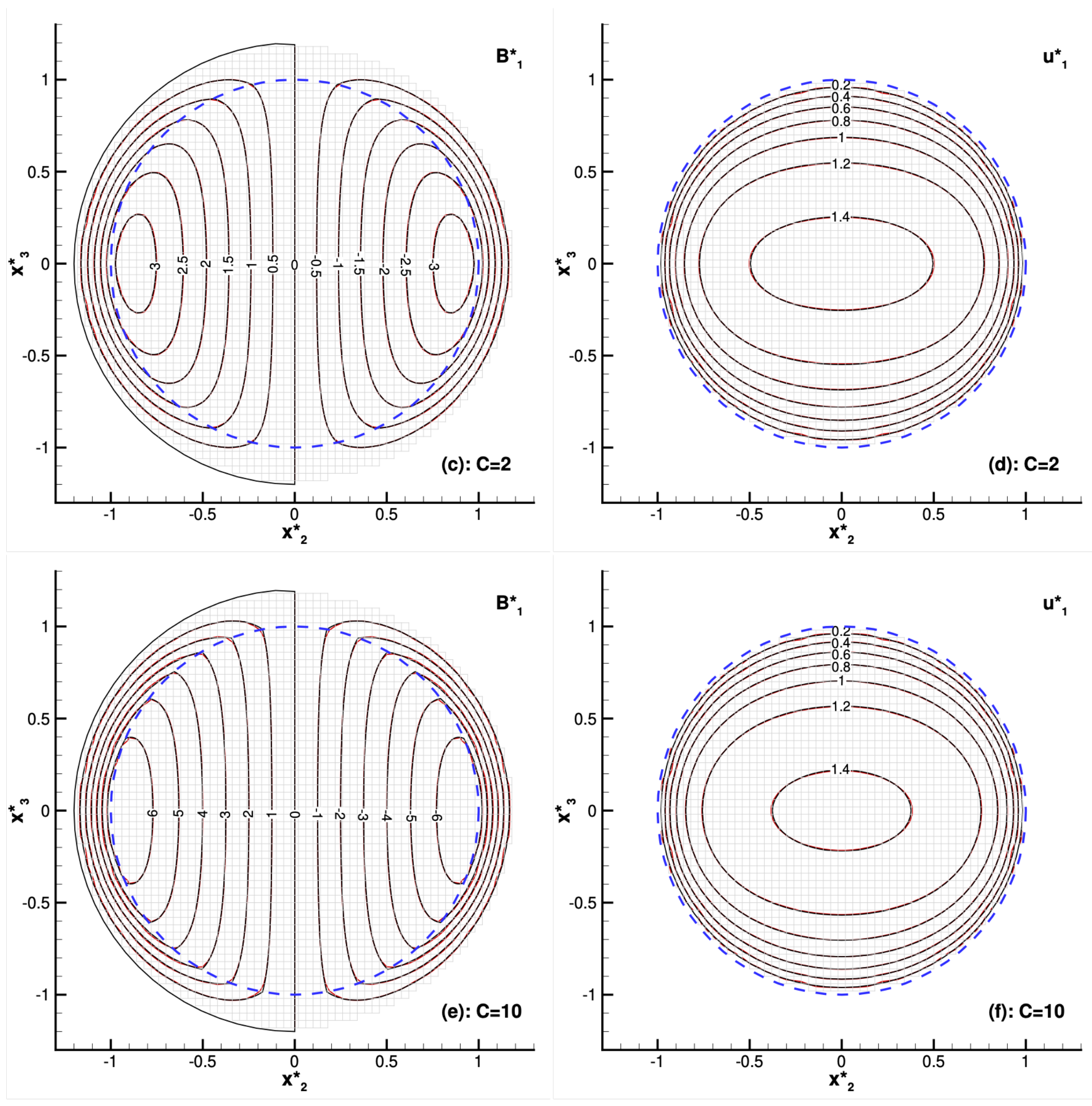

Fig. 2: Comparison between LBM results and the piecewise analytical solutions [15] in conjugate simulations of MHD pipe flows, $Ha = 10$.

## *3.2. Simulation of conjugate magnetic field with the cross-coupling mechanism*

For problems with a smooth spatial variation of $\eta$ via $\sigma$ due to temperature or composition gradients, the source term $\vec{S} = -\nabla\vec{B} \cdot \nabla\eta$ for the cross-coupling mechanism is well defined inside the whole computational domain. However, many applications have an abrupt jump in $\eta$ across the fluid-solid interface. This poses a significant challenge to numerical simulations because the physical solution becomes piecewise defined and must satisfy multiple continuity conditions at the conjugate interface, particularly in 3-D problems. Numerically, one may solve the governing equation without $\vec{S}$ separately in the fluid and solid domains, and explicitly impose these conditions at the fluid-solid interface to patch the piecewise solutions, which is difficult to implement for complex geometries. Instead, we opt to solve the governing equation (5) in the whole domain, include a non-zero source term $\vec{S}$ at grid points adjacent to the interface and compute $\vec{S}$ appropriately such that the required physical constraints at the interface are automatically satisfied with acceptable accuracy.

When $\eta$ varies with $\vec{x}$ smoothly, the source term $\vec{S}$ should be implemented using the same scheme at all grid points. Additionally, $\tau_\eta$ changes with $\eta$ according to the correlation $(\tau_\eta - 0.5)\Delta t c^2/4 = \eta$ but an error term proportional to $\nabla\tau_\eta \cdot \nabla\vec{B}$ is introduced when replacing $(1 - 0.5/\tau_\eta)\nabla \cdot [(\tau_\eta \Delta t c^2/4)\nabla\vec{B}]$ with $\nabla \cdot [(\tau_\eta - 0.5)(\Delta t c^2/4)\nabla\vec{B}]$ for $\nabla \cdot (\eta\nabla\vec{B})$ in the Chapman–Enskog analysis. This also occurs to the LBM simulations using a variable relaxation time for flow and convection-diffusion processes when the kinematic viscosity and diffusivity vary with $\vec{x}$, respectively. However, this issue is not addressed in the current study that is focused on conjugate MHD simulations where $\eta$ is constant inside the fluid and solid domains, respectively, but has an abrupt jump across the fluid-solid interface.

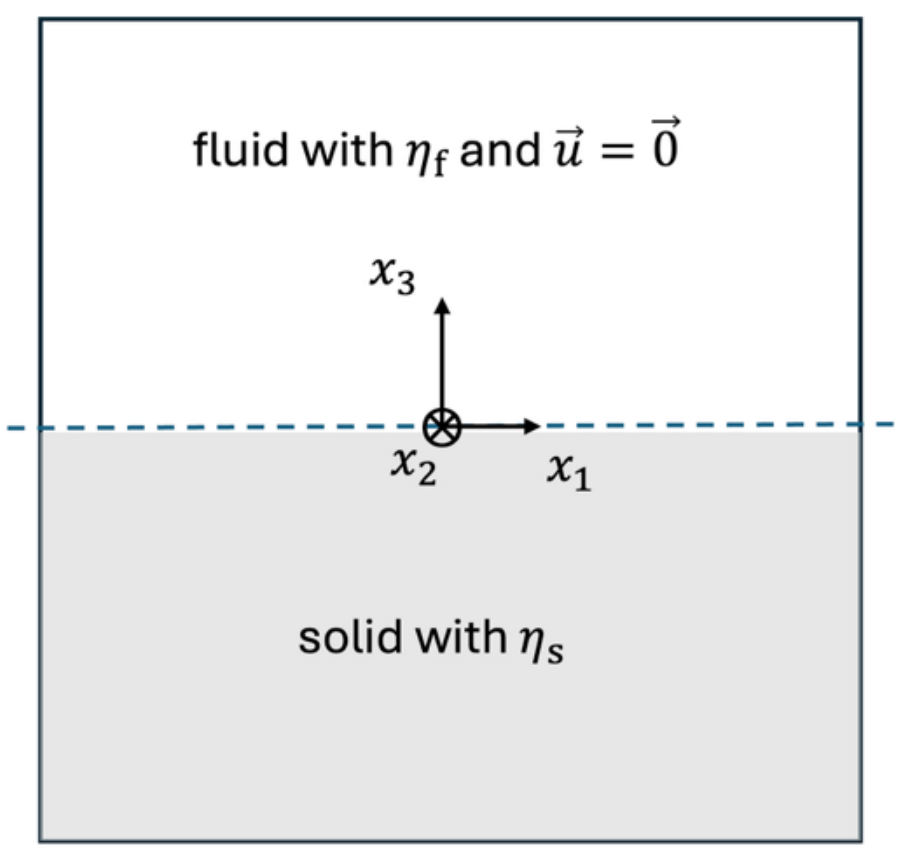

Fig. 3: Schematic for conjugate solutions of the magnetic field with a sharp interface.

A regular 3-D domain with a sharp interface at $x_3 = 0$ is shown in Fig. 3, where $\eta$ equals $\eta_\mathrm{f}$ for $x_3 \geq 0$ and $\eta_\mathrm{s}$ for $x_3 < 0$. Under the no-flow condition, a steady piecewise solution that demonstrates the influence of $\vec{S} = -\nabla\vec{B} \cdot \nabla\eta$ is constructed as follows:

$$\begin{cases} B_1 = 10 - \frac{1}{2}x_1^2 + \frac{1}{2}x_3^2 + \beta x_3, \\ B_2 = 0, \\ B_3 = 10 + x_1 x_3 + x_1, \end{cases} \tag{11}$$

where the coefficient $\beta$ equals $\beta_\mathrm{f}$ for $x_3 \geq 0$ and $\beta_\mathrm{s}$ for $x_3 < 0$ due to the jump in $\eta$, as discussed after Eq. (12). It is easy to verify that the above hypothetical analytical solution satisfies the physical continuities across the interface at $x_3 = 0$, namely (i) the continuity of $\vec{B}$ that is true, (ii) the continuity of the normal current density $J_3 = \frac{1}{\mu}\left(\frac{\partial B_2}{\partial x_1} - \frac{\partial B_1}{\partial x_2}\right)$ that is also true since $\vec{B}$ is continuous at arbitrary tangential coordinates $x_1$ and $x_2$, as well as (iii) the continuity of the tangential electric fields $E_1$ and $E_2$, which requires

$$\eta\left(\frac{\partial B_1}{\partial x_3} - \frac{\partial B_3}{\partial x_1}\right) = E_2 \quad \text{and} \quad \eta\left(\frac{\partial B_3}{\partial x_2} - \frac{\partial B_2}{\partial x_3}\right) = E_1 \tag{12}$$

to be continuous, where the former is satisfied by setting $\eta_\mathrm{f}(\beta_\mathrm{f} - 1) = \eta_\mathrm{s}(\beta_\mathrm{s} - 1)$ and the latter is always satisfied since it is zero everywhere. The required conjugate constraints at the fluid-solid interface are automatically satisfied by the improved LBM scheme, as discussed in the Appendix.

In addition to the conjugate constraints, the analytical solution also satisfies the governing equations (4) and (5) inside each subdomain. We set $\eta_\mathrm{f} = 5$, $\eta_\mathrm{s} = 2$ and $\beta_\mathrm{f} = 3$, which determine $\beta_\mathrm{s} = 6$ according to the constraint above, and use $x_1, x_3 \in [-1, 1]$ for the whole domain. An arbitrary constant of 10 is included in the analytical solution to ensure that the magnetic field components are non-negative inside the whole domain.

Note that the solution of Eq. (11) has the normal gradients of $\vec{B}$ as follows:

$$\begin{cases} \eta \dfrac{\partial B_1}{\partial x_3} = \eta(x_3 + \beta), \\ \eta \dfrac{\partial B_3}{\partial x_3} = \eta x_1, \end{cases} \tag{13}$$

which are not continuous at $x_3 = 0$ since $\eta_\mathrm{f}\beta_\mathrm{f} \neq \eta_\mathrm{s}\beta_\mathrm{s}$ and $\eta_\mathrm{f}x_1 \neq \eta_\mathrm{s}x_1$, respectively. This property is embedded in the analytical solution to expose possible errors when numerical schemes neglect the cross-coupling mechanism $\vec{S}$ and inappropriately treat $-\nabla \times (\eta\nabla \times \vec{B})$ of Eq. (3) just as a divergence term $\nabla \cdot (\eta\nabla\vec{B})$, mistakenly imposing the continuity constraint for $\eta\partial\vec{B}/\partial x_3$ at the interface with $x_3 = 0$.

The verification of the proposed LBM scheme in simulating this hypothetical problem is conducted as follows: the LBM simulation starts with an initial distribution of $\vec{B}$ estimated using $\eta_\mathrm{f} = \eta_\mathrm{s} = 2$ and $\beta_\mathrm{f} = \beta_\mathrm{s} = 3$ in the analytical solution and implements the Dirichlet boundary condition [13] according to the exact values that the analytical solution takes at the specific boundary coordinates. This might seem trivial but will verify if the LBM scheme is compatible with the analytical solution, which is true only if the LBM solution remains stabilised at the initial value. Then, we change the parameters to $\eta_\mathrm{f} = 5$ for the LBM evolution scheme inside the fluid domain and $\beta_\mathrm{s} = 6$ for the LBM boundary condition of the solid domain, which gives rise to the solution change inside both domains. After reaching convergence, we compare the steady-state LBM solution against the analytical solution.

The computational domain of $x_1, x_3 \in [-1, 1]$ is discretised using $100 \times 100$ uniform grid points with $\Delta x = 0.02$ and three grid points are used in the $x_2$ direction with periodic boundary conditions since the adopted LBM solver is developed for 3-D simulations. Note that the fluid-solid interface is located exactly at the middle between the 50th and 51st grid points in the $x_3$ direction, as required by the schematic setting; additionally, the first and the last (100th) grid points are the Dirichlet boundary locations but have the coordinates $x_3$ shifted inwards from $-1$ and $1$ by $\Delta x/2$, respectively. The distance from the Dirichlet boundary grid points to the adjacent internal grid points is always $\Delta x$, which requires extrapolation for the anti-bounceback process [13]. This grid layout ensures that the grid distance is uniform, which is required in computing $\nabla\eta$ using Eq. (8), particularly at internal grid points adjacent to boundary grid points. Similar care has been taken in the other coordinate directions.

The LBM simulation with the estimated initial setting is stable and advanced for 5000 timesteps with $\Delta t = 4 \times 10^{-5}$, after which the two parameters are changed to $\eta_\mathrm{f} = 5$ and $\beta_\mathrm{s} = 6$, respectively, to start the evolution due to the cross-coupling mechanism between different magnetic field components. The converged results obtained using the improved LBM scheme are in good agreement with the analytical solutions, as shown in Fig. 4. Additionally, the converged results obtained using the original LBM scheme, which solves $-\nabla \times (\eta\nabla \times \vec{B})$ as a divergence term $\nabla \cdot (\eta\nabla\vec{B})$, are also included for comparison. Significant errors are observed, which will further lead to noticeable errors in the coupled velocity field if the flow process is included for MHD simulations. The detailed evolution of the improved LBM simulation is presented in Fig. 5.

As shown in Fig. 4(b), the maximum discrepancy between the improved LBM results and the analytical solutions occurs in $B_3$ around the fluid-solid interface at $x_3 = 0$, which is very small (e.g., 10.57 and 10.59 at $x_1 = 0.6$ for the numerical and analytical solutions, respectively), as shown in the detailed comparison in Fig. 5(b). Therefore, the accuracy of the improved LBM scheme is good in capturing the cross-coupling mechanism between different magnetic field components due to an abrupt conductivity jump across the fluid-solid interface.

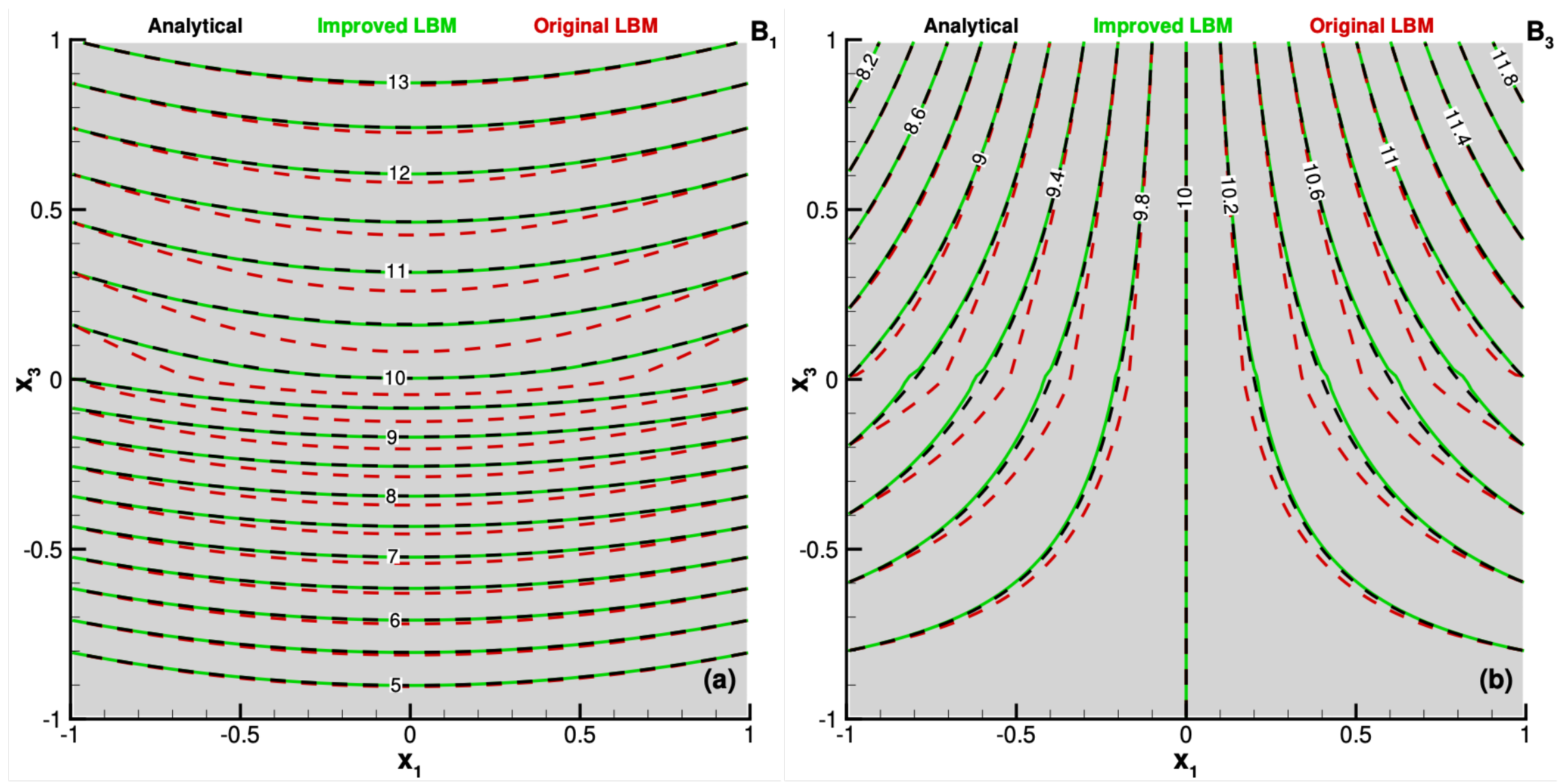

Fig. 4: Comparison of (a) $B_1$ and (b) $B_3$ between the hypothetical piecewise analytical solution of Eq. (11) with the cross-coupling mechanism and the two steady-state solutions obtained using the original LBM scheme and the improved LBM scheme of Eq. (6), respectively.

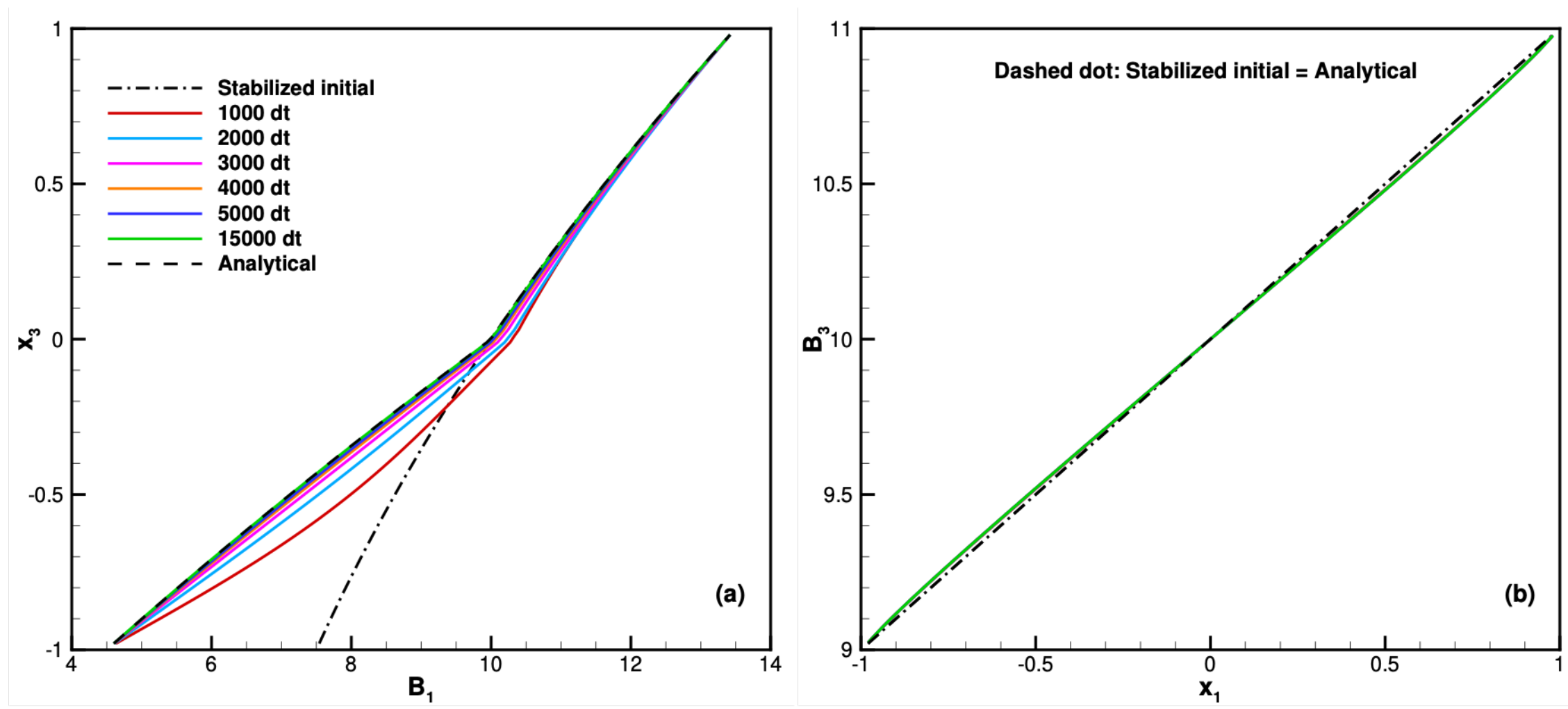

Fig. 5: Evolution of transient solutions of the improved LBM scheme of Eq. (6) towards and around the hypothetical piecewise analytical solution of Eq. (11) with the cross-coupling mechanism for (a) $B_1(x_3)$ at $x_1 = 0$ and (b) $B_3(x_1)$ at $x_3 = 0$.

## 4. Conclusions

Conjugate MHD simulations are conducted and the full magnetic induction equation is solved using an improved LBM scheme proposed in the current work. The required conjugate constraints at the fluid-solid interface are automatically satisfied by the improved LBM scheme, as discussed in the Appendix. The exact curl-of-curl term for the magnetic diffusion is considered as the sum of a divergence term and a cross-coupling term between different magnetic field components. The latter was neglected in previous MHD simulations but is required in general 3-D problems with a variable electrical conductivity. This poses a challenge in conjugate simulations involving an abrupt conductivity jump across the fluid-solid interface, which is common in practical applications. The improved LBM accurately captures this cross-coupling mechanism at the sharp interface, while the original LBM shows significant errors in comparison with a hypothetical piecewise analytical solution. The original LBM remains valid for 2-D conjugate MHD simulations since the cross-coupling mechanism vanishes automatically, as demonstrated by comparison with the piecewise analytical solution of MHD pipe flows.

## Appendix: theoretical verification of the required conjugate constraints

At the fluid-solid conjugate interface, the continuity of $\vec{B}$ is always satisfied since $\vec{B}$ is directly solved in the LBM simulations and the difference between adjacent grids are expected due to using a finite grid size. The continuity required for the normal current density is automatically satisfied due to the continuity of $\vec{B}$ at arbitrary tangential coordinates. Other conjugate constraints require continuity for the tangential electric fields, which involves normal derivatives of the components $B_i$ and can be theoretically verified for the improved LBM scheme.

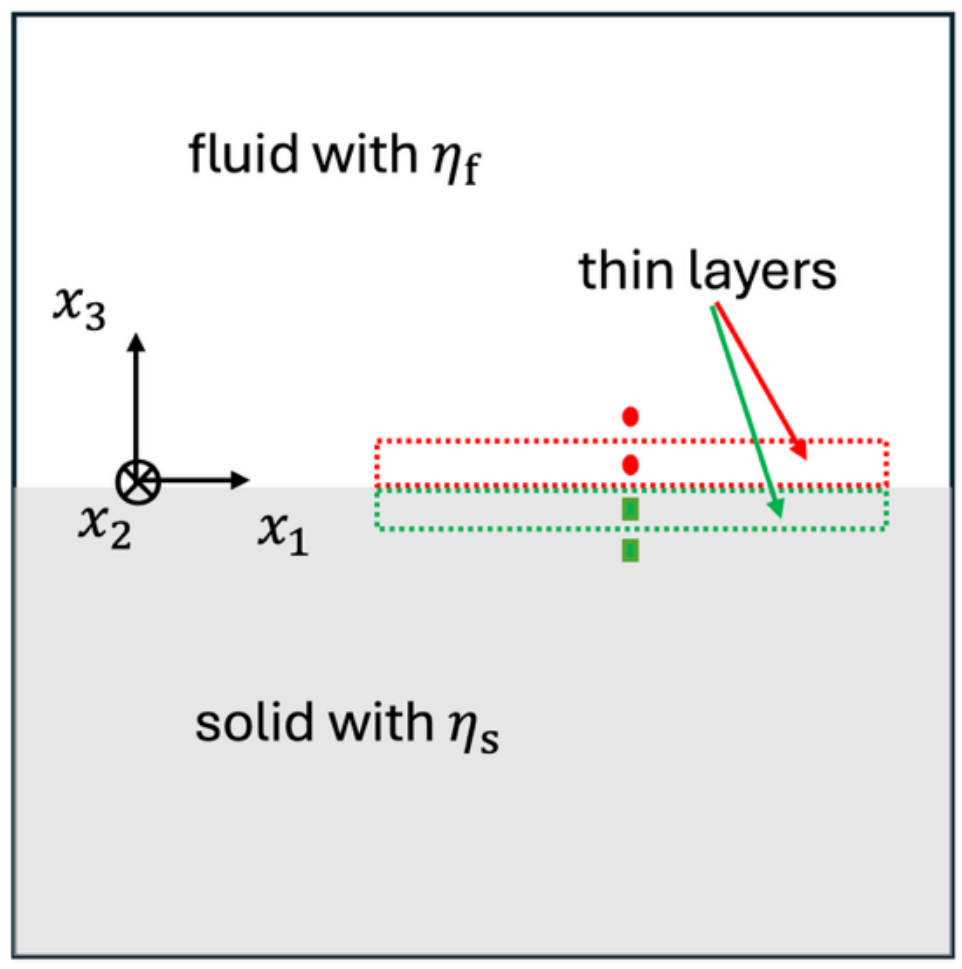

Fig. A1: Schematic for conjugate constraints imposed as a volume integral inside two adjacent boundary layers.

As shown in Fig. A1, two thin layers $\delta\Omega_f$ and $\delta\Omega_s$ adjacent to the fluid-solid interface are considered and the size of each layer is $\Delta x_1 \Delta x_2 \delta x_3$, where $\delta x_3$ is infinitesimal. The volume integral of the full magnetic induction equation (5) is as follows:

$$\iiint_{\delta\Omega_f+\delta\Omega_s}\left[-\frac{\partial\vec{B}}{\partial t}+\nabla\times\left(\vec{u}\times\vec{B}\right)+\nabla\cdot\left(\eta\nabla\vec{B}\right)-\nabla\vec{B}\cdot\nabla\eta\right]\mathrm{d}V=0, \tag{A1}$$

which can be rewritten by using Gauss's divergence theorem and considering $\delta x_3 \ll \Delta x_1, \Delta x_2$ and $\vec{u}=0$ at the stationary interface:

$$\left[\left(\eta\frac{\partial\vec{B}}{\partial x_3}\right)_{\mathrm{f}}-\left(\eta\frac{\partial\vec{B}}{\partial x_3}\right)_{\mathrm{s}}\right]\Delta x_1\Delta x_2-\iiint_{\delta\Omega_{\mathrm{f}}+\delta\Omega_s}\nabla\vec{B}\cdot\nabla\eta\,\mathrm{d}V=0, \tag{A2}$$

where the second term is negligible compared to the first term if $\eta$ is continuous for a finite $\nabla\eta$ but the two terms are comparable if $\eta$ has a jump across the interface. The subscripts f and s are used for quantities defined inside the fluid and solid layers, respectively. In LBM simulations with a grid size equal to $\delta x_3$, Eq. (8) is adopted to compute $\nabla\eta$ at each fluid or solid grid point using values of $\eta$ at the neighbouring grid points, having:

$$\frac{\partial\eta}{\partial x_3}=\frac{1}{2\delta x_3}(\eta_{\mathrm{f}}-\eta_s), \tag{A3}$$

which holds inside both layers at the red dot and green square in Fig. (A1). Note that the factor of 2 in the denominator is required for the following analysis and Eq. (A*3*) is also valid for usual finite difference schemes with a second-order accuracy. Meanwhile, the spatial derivatives of $B_i$ are defined using the piecewise solutions inside the fluid and solid domains, respectively, which is satisfied in LBM simulations by using the local summation scheme of Eq. (9) at each grid point of each domain piece. Therefore, Eq. (A2) becomes:

$$\left[\left(\eta\frac{\partial\vec{B}}{\partial x_3}\right)_{\mathrm{f}}-\left(\eta\frac{\partial\vec{B}}{\partial x_3}\right)_{\mathrm{s}}\right]-\frac{1}{2}(\eta_{\mathrm{f}}-\eta_s)\left[\left(\frac{\partial B_3}{\partial\vec{x}}\right)_{\mathrm{f}}+\left(\frac{\partial B_3}{\partial\vec{x}}\right)_{\mathrm{s}}\right]=0. \tag{A4}$$

The first component of Eq. (A4) is:

$$\left[\left(\eta\frac{\partial B_1}{\partial x_3}\right)_{\mathrm{f}}-\left(\eta\frac{\partial B_1}{\partial x_3}\right)_{\mathrm{s}}\right]-\frac{1}{2}(\eta_{\mathrm{f}}-\eta_s)\left[\left(\frac{\partial B_3}{\partial x_1}\right)_{\mathrm{f}}+\left(\frac{\partial B_3}{\partial x_1}\right)_{\mathrm{s}}\right]=0, \tag{A5}$$

which, by using the continuity of $B_i$ across the interface (equivalent to the continuity of $\partial B_i/\partial x_1$ and $\partial B_i/\partial x_2$ when observed along $x_1$ and $x_2$, respectively), can be rewritten as:

$$\left(\eta\frac{\partial B_1}{\partial x_3}\right)_{\mathrm{f}}-\frac{1}{2}\eta_{\mathrm{f}}\left[\left(\frac{\partial B_3}{\partial x_1}\right)_{\mathrm{f}}+\left(\frac{\partial B_3}{\partial x_1}\right)_{\mathrm{f}}\right]=\left(\eta\frac{\partial B_1}{\partial x_3}\right)_{\mathrm{s}}-\frac{1}{2}\eta_s\left[\left(\frac{\partial B_3}{\partial x_1}\right)_{\mathrm{s}}+\left(\frac{\partial B_3}{\partial x_1}\right)_{\mathrm{s}}\right], \tag{A6}$$

which implies the continuity of the tangential electric field $E_2$ across the interface, as shown in Eq. (12). Similarly, it is easy to verify that the second component of Eq. (A4) implies the continuity of the tangential electric field $E_1$, as also shown in Eq. (12). The third component of Eq. (A4) is:

$$\left[\left(\eta\frac{\partial B_3}{\partial x_3}\right)_{\mathrm{f}}-\left(\eta\frac{\partial B_3}{\partial x_3}\right)_{\mathrm{s}}\right]-\frac{1}{2}(\eta_{\mathrm{f}}-\eta_s)\left[\left(\frac{\partial B_3}{\partial x_3}\right)_{\mathrm{f}}+\left(\frac{\partial B_3}{\partial x_3}\right)_{\mathrm{s}}\right]=0, \tag{A7}$$

which, after removing the constant $(\eta_{\mathrm{f}}+\eta_s)/2$, can be rewritten as:

$$\left(\frac{\partial B_3}{\partial x_3}\right)_{\mathrm{f}}=\left(\frac{\partial B_3}{\partial x_3}\right)_{\mathrm{s}}, \tag{A8}$$

which is consistent with the divergence-free condition of $\nabla\cdot\vec{B}=0$ in both layers after considering the continuity of $\partial B_1/\partial x_1+\partial B_2/\partial x_2$ across the interface. Therefore, the improved LBM scheme imposes two conjugate constraints for the tangential electric fields and reinforces the divergence-free condition inside bulk areas around the interface.

## Data availability

Data will be made available on request.

## Declaration of competing interest

The authors declare that they have no known competing financial interests or personal relationships that could have appeared to influence the work reported in this paper.